# Passive and active suppression of transduced noise in silicon spin qubits

Jaemin Park[1†], Hyeongyu Jang[1†], Hanseo Sohn[1], Jonginn Yun[1], Younguk Song[1], Byungwoo Kang[1], Lucas E. A. Stehouwer[2], Davide Degli Esposti[2], Giordano Scappucci[2], and Dohun Kim[1]*

[1]*Department of Physics and Astronomy, and Institute of Applied Physics, Seoul National University, Seoul 08826, Korea*

[2] *QuTech and Kavli Institute of Nanoscience, Delft University of Technology, PO Box 5046, 2600 GA Delft, The Netherlands*

[†]*These authors contributed equally to this work.*

*Corresponding author: dohunkim@snu.ac.kr*

## Abstract

Addressing and mitigating decoherence sources plays an essential role in the development of a scalable quantum computing system, which requires low gate errors to be consistently maintained throughout the circuit execution. While nuclear spin-free materials, such as isotopically purified silicon, exhibit intrinsically promising coherence properties for electron spin qubits, the omnipresent charge noise, when converted to magnetic noise under a strong magnetic field gradient, often hinders stable qubit operation within a time frame comparable to the data acquisition time. Here, we demonstrate both open- and closed-loop suppression techniques for the transduced noise in silicon spin qubits, resulting in a more than two-fold (ten-fold) improvement of the inhomogeneous coherence time (Rabi oscillation quality) that leads to a single-qubit gate fidelity of over 99.6% even in the presence of a strong decoherence field gradient. Utilizing gate set tomography, we show that adaptive qubit control also reduces the non-Markovian noise in the system, which validates the stability of the gate fidelity. The technique can be used to learn multiple Hamiltonian parameters and is useful for the intermittent calibration of the circuit parameters with affordable experimental overhead, providing a useful subroutine during the repeated execution of general quantum circuits.

## Introduction

Spins in semiconductor quantum dots (QDs) offer a promising platform for developing large-scale quantum information processors[1–3]. Benefitting from isotopic purification, silicon spin qubits have recently enabled compelling demonstrations of coherent quantum operations of multiple qubits[4–7]. Single and two-qubit gate fidelities exceeding 99%[8–11] were realized along with elementary quantum error correction[12]. However, for fault-tolerant quantum computing, achieving consistent high-fidelity single- and two-qubit control is crucial, not only over an extended time but also across an extensive array of devices[13,14]. Therefore, the ability to precisely control the qubit parameters and a deeper understanding of the origin of the noise affecting the qubit system are pivotal steps toward developing mitigation strategies.

Real-time adaptive control, a powerful tool for stabilizing qubit operation via the active suppression of noise[15–19], allows the precise manipulation of quantum states. Experimental validations have been conducted, for example, in superconducting qubits[20,21], spins in diamond[22], and trapped atoms[23]. In the silicon QD platform, previous studies to address unwanted interactions between the spin qubits and the environment have focused on the automated correction of slowly drifting system parameters. These studies demonstrated, for example, reliable single- and two-qubit parameter calibration, including optimal readout points, using a field programmable gate array (FPGA)[7,24]. Although rapid progress has been made in optimizing the electrical controllability[25,26] while benefitting from intrinsically low magnetic noise sources, many of the silicon spin qubits fabricated thus far are negatively affected by transduced noise, which often extends to a wide range of frequencies. Dynamical decoupling methods such as spin-echo can extend the coherence time when the qubits are idle, but incorporating these complex pulse sequences with a general quantum algorithm is not straightforward. On the other hand, rapid estimation of the Hamiltonian parameter by Bayesian inference[27] has the potential for fast parameter calibration and is compatible with arbitrary qubit operations. Thus, the method can be used to improve the performance of qubits in a wide variety of quantum information processing applications.

Here, we demonstrate the rapid and real-time noise suppression in a $^{28}$Si/SiGe spin qubit device. We focus on both open- and closed-loop control on silicon spin qubits where the typical time scale of parameter fluctuation, due mainly to charge noise transduced to magnetic noise in the presence of a micromagnet, is comparable to the data acquisition time. For open-loop noise suppression, we investigate the controllable backaction of the charge sensor to the qubit where the rf-single electron transistor (rf-SET), referred to as a sensor dot (SD), close to the qubit acts as a noise source. In this case, the dynamic pulsing of the sensor chemical potential to the Coulomb blockade regime is used to suppress the noise in the qubit manipulation phase.

The remaining noise is additionally suppressed by a hardware-implemented Bayesian inference and frequency feedback circuit. By applying a methodology similar to that successfully demonstrated in GaAs devices[15–19], we confirm that fast parameter estimation and adaptive control can also boost the performance of qubits in $^{28}$Si/SiGe devices. The confirmed results reveal the general applicability of the active suppression technique based on Hamiltonian parameter estimation for both charge noise and nuclear spin noise suppression. Compared with the bare single-qubit gate fidelity of less than 98.6%, the technique enables single-qubit gate fidelity above 99.6% even in the presence of a large local magnetic field gradient and a significant source of charge noise. We further confirm the stability of the single-qubit gate performance using gate set tomography (GST) and discuss the potential of applying the method to fast two-qubit parameter estimation.

## Results

### The charge sensor-spin qubit system.

We fabricated QD array devices with an overlapping gate layout[7,28] on an isotopically purified $^{28}$Si/SiGe heterostructure wafer (refer to the Methods section for fabrication details). The device is composed of an array of five gate-defined QDs and two SDs on both sides, as illustrated in Fig. 1a. We focus on the operation and measurement of the left SD and the leftmost qubit. On top of the device, we deposited a cobalt-based micromagnet of which the geometry is similar to that in previous studies[26]. Figure 1b shows the simulated magnetic field distribution near the qubit array, offset by an applied homogeneous magnetic field of 440 mT. In the line cut of the field profile, as shown in Fig. 1c, the measured qubit frequency is in good agreement with the simulation, revealing a strong spatial gradient $dB_z/dx = 0.184$ mT/nm at the location of the qubit. This enables each qubit to be individually and electrically addressed but simultaneously acts as a decoherence source.

Qubit manipulation involves applying a burst of microwaves to the upper screening gate $V_{\mathrm{screen}}$ to induce electric dipole spin resonance (EDSR). We use an energy-selective tunneling process near the charge transition of the last electron for qubit initialization and readout[29]. The SD is connected to an LC tank circuit for rf-reflectometry[30–32], which is performed by injecting a carrier signal at the frequency of 143 MHz and power of –100 dBm. The reflected power is monitored through cryogenic and room temperature amplification and subsequent homodyne detection with an integration time of 2 μs. See Supplementary Note. 1 for details of measurement setup.

**Passive noise suppression.**

We first demonstrate passive suppression of the transduced noise. Figure 2a illustrates the qubit capacitively coupled to SD. The backaction of the SD on the qubit[33,34] is marked by arrows in Fig. 2a, which arises from electron transport through the Coulomb blockade-lifted SD, whose fluctuations lead to that of the position of the qubit. This type of noise was commonly investigated by detecting the discrete fluctuations in the transport current[31], but here we investigate its effect on the qubit coherence. Although it is commonly believed that this backaction can be lowered by turning off the carrier power for the qubit manipulation phase[35,36], we note that significant qubit dephasing still occurs as long as electron transport is allowed in the SD (see Supplementary Note. 2 for details of sensor-qubit coupling). Thus, the main way to suppress the noise is to dynamically pulse the plunger gate of the SD, $V_{\mathrm{SET}}$, by the amount of $\varepsilon$ towards the Coulomb blockaded regime during qubit manipulation, as schematically shown in Fig. 2b. For the qubit measurement, $V_{\mathrm{SET}}$ is pulsed back to the regime in which the sensor is maximally sensitive to changes in the charge number in the nearby QD.

Figure 2c shows the variation in the inhomogeneous coherence time $T_2^*$, measured by Ramsey interference measurement, and Rabi decay time $T_2^{\mathrm{Rabi}}$ as functions of $\varepsilon$. Starting from the minimum $T_2^*$ ($T_2^{\mathrm{Rabi}}$) of about 0.92 (2.52) μs when maximum transport current is allowed through SD ($\varepsilon = 6$ mV), the qubit demonstrates improved coherence as the chemical potential of SD is pulsed towards the Coulomb blockade regime, and the behavior is expected to be periodic per that of the Coulomb oscillations in SD. For $\varepsilon = -6$ mV, we observe an increase in $T_2$* ($T_2^{\mathrm{Rabi}}$) of more than 75% (500%) compared with the case of $\varepsilon = 0$ mV. Consequently, as shown in Fig. 2d, the Rabi oscillation quality under resonant conditions is significantly increased, thereby attesting to the effectiveness of the dynamic pulsing of the SD chemical potential. Moreover, taking into account frequency detuning, a comparison of the Rabi chevron pattern between the case of $\varepsilon = 0$ mV (Fig. 2e) and $\varepsilon = -6$ mV (Fig. 2f) also shows that the qubit driving quality is directly enhanced, as Fig. 2f exhibits improved clarity and contrast in probability oscillations. The resonant frequency shift in Fig. 2f, which corresponds to pulsing by $\varepsilon = -6$ mV, using the field distribution determined in Fig. 1c, translates to a position shift of the qubit by about 0.33 nm, which is unlikely to lead to significantly different distributions of the nuclear spins in the host material. Therefore, the decoherence is primarily attributed to the charge noise transduced to magnetic noise.

**Noise spectroscopy.**

Fluctuations in the qubit frequency were observed in both the time and frequency domains through repeated Ramsey experiments for $\varepsilon = 0$ and –6 mV, as shown in Fig. 3a and 3b. Comparing Fig. 3a and 3b, we note that slow frequency fluctuation remains even after the application of passive noise suppression. Although suppressing the noise originating from the SD notably enhanced $T_2^*$ and $T_2^{\mathrm{Rabi}}$, the effect is not apparent in the repeated Ramsey experiment, due to the limited frequency sampling rate in this case of 0.548 Sa/s.

To enhance the frequency estimation rate, we employed Bayesian inference of the qubit frequency based on one hundred single-shot outcomes of the Ramsey experiment where the free evolution time increased from 40 ns to 4 μs in steps of 40 ns (refer to the Methods section for details). Consequently, the estimation cycle takes 24 ms (100 × 240 μs), where one shot of measurements consists of a microwave burst and waiting time of 60 μs for reducing the heating effect[7,37], a readout duration of 140 μs, and a calculation time of 40 μs for Bayesian inference.

Figure 3c shows the variance of the frequency fluctuation $\sigma^2 = 2D_\alpha T^\alpha$ as a function of time interval $T$ where $D_\alpha$ is the diffusion coefficient. We note that $\sigma^2$ follows sub-diffusive behavior[38] with an exponent $\alpha = 0.58$ (0.47) and $D_\alpha = 0.0179$ MHz$^2$/s$^{0.58}$ (0.0119 MHz$^2$/s$^{0.47}$) for $\varepsilon = 0$ mV (–6 mV). The reduction of $\alpha$ and $D_\alpha$ suggests the suppression of noise stemming from the SD backaction over $T$. Using 30,000 samples obtained from the Bayesian inference, we also observe that dynamically pulsing SD results in an overall reduction in the noise power spectral density PSD, as shown in Fig. 3d. We fit the PSD to power-law spectra $A/f^\beta$ with a noise amplitude $A$ = 0.00296 MHz$^2$/Hz (0.00175 MHz$^2$/Hz) and an exponent $\beta = 1.34$ (1.17) for $\varepsilon = 0$ mV (–6 mV). Thus, we confirm that, as the sensor backaction is suppressed by dynamic pulsing, so is the noise amplitude of fluctuations in the qubit frequency.

To further analyze the PSD, we focus on the decay envelope of Ramsey oscillation with the free evolution time $t$, as represented by the decoherence function (see Supplementary Note. 3 for details of derivation)[39,40]

$$W(t) = \exp\left(-t^2 \Big/ \left(T_2^*\right)^2\right) = \exp\left(-\frac{t^2}{2}(2\pi)^2 \int_{f_0}^{\infty} df\, S(f)\,\mathrm{sinc}^2(\pi f t)\right) \quad (1)$$

where $f_0$ is the reciprocal of the total integration time of the experiment (~ 5 min) and $S(f)$ is the PSD. Substituting the fitting parameters $A$ and $\beta$ in $S(f)$, the integral yields theoretical $T_2$* of 0.916 μs for $\varepsilon = 0$ mV and 1.404 μs for $\varepsilon = -6$ mV, which are in good agreement with the experimentally measured values (Fig. 2c) within the uncertainty of the fitting procedures. Moreover, one can obtain the quasi-static variance $\sigma_{\mathrm{static}} = 1/\sqrt{2}\pi T_2^*$ from Eqn. (1) of 245.69 kHz (160.28 kHz) for $\varepsilon = 0$ mV (–6 mV). $\Delta\sigma_{\text{static}}$ of 85.41 kHz reflects the reduced noise achieved by dynamic SD pulsing while residual noise 160.28 kHz may stem from other transduced noise sources distributed in the system that are not properly controlled.

**Active noise suppression.**

Applying the passive noise suppression technique by default, we further implemented active feedback control to suppress the remaining noise. We used the Bayesian inference described above as a probing tool and separated the experimental sequence into probe and operation steps, as illustrated in Fig. 4a. To collect 100 single-shot outcomes during the probe phase, $\pi/2$ rotation pulses separated by varying free evolution time were applied, where the frequencies $f_{\text{MW}}$ are detuned by 2 MHz from the local oscillator frequency. We then calculated the instantaneous estimation of the frequency $f_{\text{est}}$ and the frequency error $\Delta f$ between the target frequency $f_{\text{target}}$ and $f_{\text{est}}$. Subsequently, the microwave frequency $f_{\text{MW}}$ was adaptively adjusted by $\Delta f$ in the operation step.

Figure 4b demonstrates the stabilization of the qubit frequency during the operation phase by successfully locking it to $f_{\text{target}} = 2$ MHz. The inset in Fig. 4b displays a histogram of qubit frequencies with (red) and without (blue) adaptive control. As $f_{\text{MW}}$ is adaptively corrected, the histogram exhibits a narrow distribution with a frequency uncertainty of 65.3 kHz, reflecting a decrease in deviation of 40.7% in comparison with $\sigma_{\text{static}} = 160.28$ kHz that was obtained from the noise spectrum. Furthermore, as shown in Fig. 4c, the noise amplitude of PSD decreases from 1553 to 837 $\text{kHz}^2/\text{Hz}$, indicating the suppression of residual noise. The flattened low-frequency noise of about $2.5\times10^{10}$ $\text{Hz}^2/\text{Hz}$ below the bandwidth of 0.1 Hz is indicative of successful noise filtering by active feedback control.

The stabilized qubit operation is also evident from the repeated Ramsey experiment with active frequency feedback, as shown in Fig. 4d. As shown in Fig. 4e, the Ramsey oscillations with active feedback reveal a more than 2-fold improvement in $T_2$* of 3.21 μs. This improvement corresponds to a deviation of 70.11 kHz estimated from $\sigma_{\mathrm{static}} = 1/\sqrt{2}\pi T_2^*$, which is also close to the frequency uncertainty in the histogram in Fig. 4b. The fitted red solid line in Fig. 4e also yields an oscillation frequency of 1.93 MHz, in excellent agreement with the target frequency of 2 MHz within the uncertainty of the frequency estimation of 70.11 kHz.

As an example of the compatibility of the method with general operation sequences, we perform a Rabi chevron experiment with controlled detuning by adaptive control. Fig. 4f shows a clear improvement of oscillation quality and contrast in probability oscillation compared with that of non-adaptive control (Fig. 2f). Different from the passive suppression technique, active feedback control enables effective noise suppression without requiring precise knowledge of the noise origin. This universality has been demonstrated across the various platforms for quantum information processing, as discussed in the introduction above, and implies scalability towards multi-qubit correction. In Supplementary Note 4. we present preliminary data showing qubit frequency estimation when strong exchange interaction is present in the two-qubit system, which further shows the possibility of applying the developed method to correct multiple Hamiltonian parameters[41,42].

## Gate set tomography

We turn to confirm the improved stability of the qubit operation enabled by adaptive control using GST[43,44], which allows the detection of specific errors for each circuit. We perform a single-qubit GST combined with frequency feedback of which schematic pulse sequence is shown in Fig. 5a. This combined GST protocol extends the total experiment time to over 6 hours, allowing a thorough assessment of the robustness of the entire system. The density matrix showing initialization and measurement fidelity (top row) and the Pauli transfer matrix for gates X/2 and Y/2 (bottom row) are shown in Fig. 5b (Fig. 5c) for GST without (with) applying passive and active noise suppression techniques. The initialization and measurement fidelity remain unchanged, indicating that the developed methods do not address state preparation and measurement (SPAM) errors. Instead, these techniques enhance the gate fidelities, $F_{X/2}$ from 98.56% to 99.66% and $F_{Y/2}$ from 98.57% to 99.49%, showing the improved stability of the qubit system.

The GST protocol is based on a Markovian gate set, assuming stationary noises for error prediction. Consequently, the GST model fails to accurately fit data influenced by non-Markovian noise the degree of which can be evaluated by the goodness-of-fit[43,44]. The method determines that the noise is sufficiently Markovian if the following inequality is satisfied.

$$k-\sqrt{2k} < 2\Delta\log L_s < k+\sqrt{2k} \quad (2)$$

where $\log L_s$ is the log-likelihood ratio between the predicted and observed probability and $2\Delta\log L_s = 2\log L_{\max,s} - 2\log L_s$ is expected to follow the $\chi_k^2$ distribution with a mean $k$ and standard deviation $\sqrt{2k}$ if the observed data is well fitted to the model. $\log L_{\max,s}$ shows the theoretical upper bound of the GST model, and $k$ is the number of independent outcomes of a single circuit[44].

Figure. 5d shows a significant degree of non-Markovian noise in the non-adaptive control case indicated by the colored boxes, displaying the values $2\Delta\log L_s$ that are outside of the range in Eqn. (2) (with a confidence level of 95%, 17 < $2\Delta\log L_s$ ). In Fig. 5e, the use of both active and passive techniques significantly reduces the impact of non-Markovian noise, as indicated by the decrease in the total amount of the log-likelihood ratios $2\Delta\log L$ = $\sum_s 2\Delta\log L_s$ from 475.7, 3122.8, 4805.3, 6169.1, and 8445.5 (Fig. 5d) to 470.9, 741, 2295.4, 3386.2, and 4835 (Fig. 5e) at maximum lengths of 1, 2, 4, 8, and 16, respectively.

## Discussion

We note that, as shown in Fig. 5e, the violated $2\Delta\log L_s$ that appear differently for each circuit, especially where the Gx germ is used at the length of 16, may indicate temporally inhomogeneous and stochastic errors, which could not be addressed by either technique, such as slow fluctuations in the readout point. Although the magnitude of the goodness-of-fit quantified by GST does not correspond to the actual amplitude of non-Markovian noise, the model violations indicate that the system is unlikely to adhere to a Markov process with high probability. These non-Markovian properties were evident from the noise spectroscopy exhibiting sub-diffusive behavior attributed to time-correlation as depicted by the nonlinear function in Fig. 3c. In this context, the Bayesian inference employed to reveal the noise characteristics also enabled us to suppress the corresponding noise.

Of the two noise suppression techniques developed in this work, the passive suppression technique by dynamically pulsing SD provides a simple mitigation strategy to minimize the effect of the detector while maintaining high charge sensitivity benefiting from strong sensor-qubit capacitive coupling. Moreover, using SD as a controllable noise source combined with fast noise spectroscopy (see Supplementary Note. 5) may offer a deeper understanding of transduced noise characteristics[45]. The noise source-agnostic active feedback strategy provides a general noise suppression method with an affordable experimental overhead.

In the future, a higher sampling rate for noise estimation will enable noise analysis in the wider frequency domain and further decrease the gate infidelity which can be realized for example by optimizing the real-time Bayesian calculation time on the hardware level, minimizing the heating effect that enables minimal waiting time after qubit manipulation, and by reducing the readout time using different spin to charge conversion method such as Pauli spin blockade-based parity measurements[7,46] (see Supplementary Note 6). Parallel with efforts to reduce charge noise source by material development[47–49] and optimizing device fabrication steps[50–52], fast intermittent correction of qubit parameter fluctuation[19,53–55] can further enhance the performance of the quantum measurement system.

## Methods

### Device fabrication

Quantum dot qubit devices were fabricated on an undoped $^{28}$Si/SiGe heterostructure featuring a 9 nm quantum well with a residual $^{29}$Si concentration of 0.08%. The quantum well is grown on a strain-relaxed $Si_{0.7}Ge_{0.3}$ substrate and is separated from the surface by a 30 nm $Si_{0.7}Ge_{0.3}$ spacer terminated by an amorphous Si-rich layer[48]. After defining the active region and alignment markers by reactive ion etching, an ohmic region was created via ion implantation with phosphorus, and the device was subsequently annealed to activate the implanted carriers. To suppress leakage, a 30 nm layer of $Al_2O_3$ is deposited on the substrate using Atomic Layer Deposition (ALD). Metal gates and additional oxide layers are formed through repetitive electron-beam lithography steps, metal evaporations of 5 nm Ti / 30 nm Pd using an e-beam evaporator, and ALD steps. The patterned micromagnet on top of the final $Al_2O_3$ layer was deposited using an e-beam evaporator.

**Bayesian inference**

Qubit frequency estimation was conducted using Bayes' rule[15,27]. The estimation was based on measurement information $m_k$ obtained from the single-shot outcome of the Ramsey oscillation experiment with a free evolution time $t_k$ = $k$ × 40 ns. The posterior distribution was approximated using an oscillatory likelihood function,

$$P(f \mid m_{\mathrm{N}}, m_{\mathrm{N}-1}, \ldots m_1) = P_0(f) \prod_{k=1}^{\mathrm{N}} \frac{1}{2} [1 + r_k(\alpha + \beta \cos(2\pi f t_k + \theta))] \quad (1)$$

where we used the repetition number N = 100 per one frequency estimation. $r_k$ = 1 (-1) for $m_k = \downarrow (\uparrow)$, $\theta$ denotes the initial phase of off-resonant Ramsey oscillation, and $\alpha$ ($\beta$) is determined by errors in the axis of rotation on the Bloch sphere (oscillation visibility). For initial frequency estimation, $P_0(f)$ was initialized as a uniform distribution reflecting a lack of prior knowledge and multiplied by the likelihood functions with $r_1$ and $t_1$. After the N$^{\mathrm{th}}$ single-shot measurement and update, the most probable $f_{\mathrm{est}}$ was determined from $P(f \mid m_{100}, m_{99}, \ldots m_1)$. However, whenever the previously estimated value of $f_{\mathrm{est}}$ is available, we set a Gaussian prior distribution $P_0(f)$ with a mean equal to the previously estimated value having a constant deviation of 50 kHz.

**Gate Set Tomography (GST)**

GST characterizes the behavior of quantum gates, the building blocks of quantum circuits, and provides the dataset for analyzing how well these gates perform their intended operations. For the specific purpose of identifying errors, GST uses structured and periodic sequences. Two fiducials {null, X/2, Y/2, X/2∘X/2, X/2∘X/2∘X/2, Y/2∘Y/2∘Y/2} constitute a small 6 × 6 circuit structure as shown in Fig. 5d and 5e. The germs {I, X/2, Y/2, X/2∘Y/2, X/2∘X/2∘Y/2} between the two fiducials, as shown in Fig. 5a, amplify errors by repeating gate operations. The datasets are generated by recording the results for each specific circuit.

**Fidelity for gate set tomography**

In GST, a quantum state is typically described by a $d \times d$ density matrix $\rho$ on a $d$-dimensional Hilbert space, while superoperators $M$ are represented as $d^2 \times d^2$ matrices in the Hilbert-Schmidt space where a basis is chosen as the four Pauli matrices $\{\sigma_I, \sigma_X, \sigma_Y, \sigma_Z\}$ for $d = 2$. GST estimates of $\rho_{\text{exp}}$ and $M_{\text{exp}}$ involve using a maximum-likelihood estimator to compute completely positive and trace-preserving (CPTP) matrices, given raw measurement outcomes for each circuit in the dataset. Initialization and measurement fidelities can be calculated by evaluating the closeness between the ideal quantum state $\rho_{\text{ideal}}$ and the measured state $\rho_{\text{exp}}$, $F_{I,M} = \left(\mathrm{Tr}\sqrt{\sqrt{\rho_{\text{ideal}}}\rho_{\text{exp}}\sqrt{\rho_{\text{ideal}}}}\right)^2$. Furthermore, by comparing the measured Pauli transfer matrix $M_{\text{exp}}$ with the ideal $M_{\text{ideal}}$, the gate fidelity is given by $F_{\text{gate}} = \left(\mathrm{Tr}\left(M_{\text{exp}}^{-1} M_{\text{ideal}}\right) + d\right) \big/ [d(d+1)]$.

Error analysis was conducted using Monte Carlo bootstrap resampling. We assume sampling distributions for the CPTP matrices obtained from the GST estimates, and then random samples were repeatedly drawn from these distributions. We generate 1,000 datasets to analyze, producing point estimates for each, and use their deviation to define a confidence region.

**Data availability**

The data that support the findings of this study are available from the corresponding author upon request.

**Acknowledgments**

This work was supported by a National Research Foundation of Korea (NRF) grant funded by the Korean Government (MSIT) (No. 2019M3E4A1080144, No. 2019M3E4A1080145, No. 2019R1A5A1027055, RS-2023-00283291, RS-2024-00413957, RS-2024-00442994, SRC Center for Quantum Coherence in Condensed Matter RS-2023-00207732, and No.

2023R1A2C2005809) and a core center program grant funded by the Ministry of Education (No. 2021R1A6C101B418). Lucas E. A. Stehouwer and Davide Degli Esposti developed and characterized the heterostructure under the supervision of Giordano Scappucci. Correspondence and requests for materials should be addressed to DK (dohunkim@snu.ac.kr).

## Author contributions

J.P. fabricated the device with the help of B.K., H.S., J.Y., and Y.S. J.P. and H.J. performed the measurements and configured the measurement software. L.E.A.S., D.D.E., and G.S. synthesized and provided the $^{28}$Si/SiGe heterostructure. All authors contributed to the preparation of the manuscript. D.K. supervised the project.

## Competing interests

The authors declare no competing interests.

## Figures

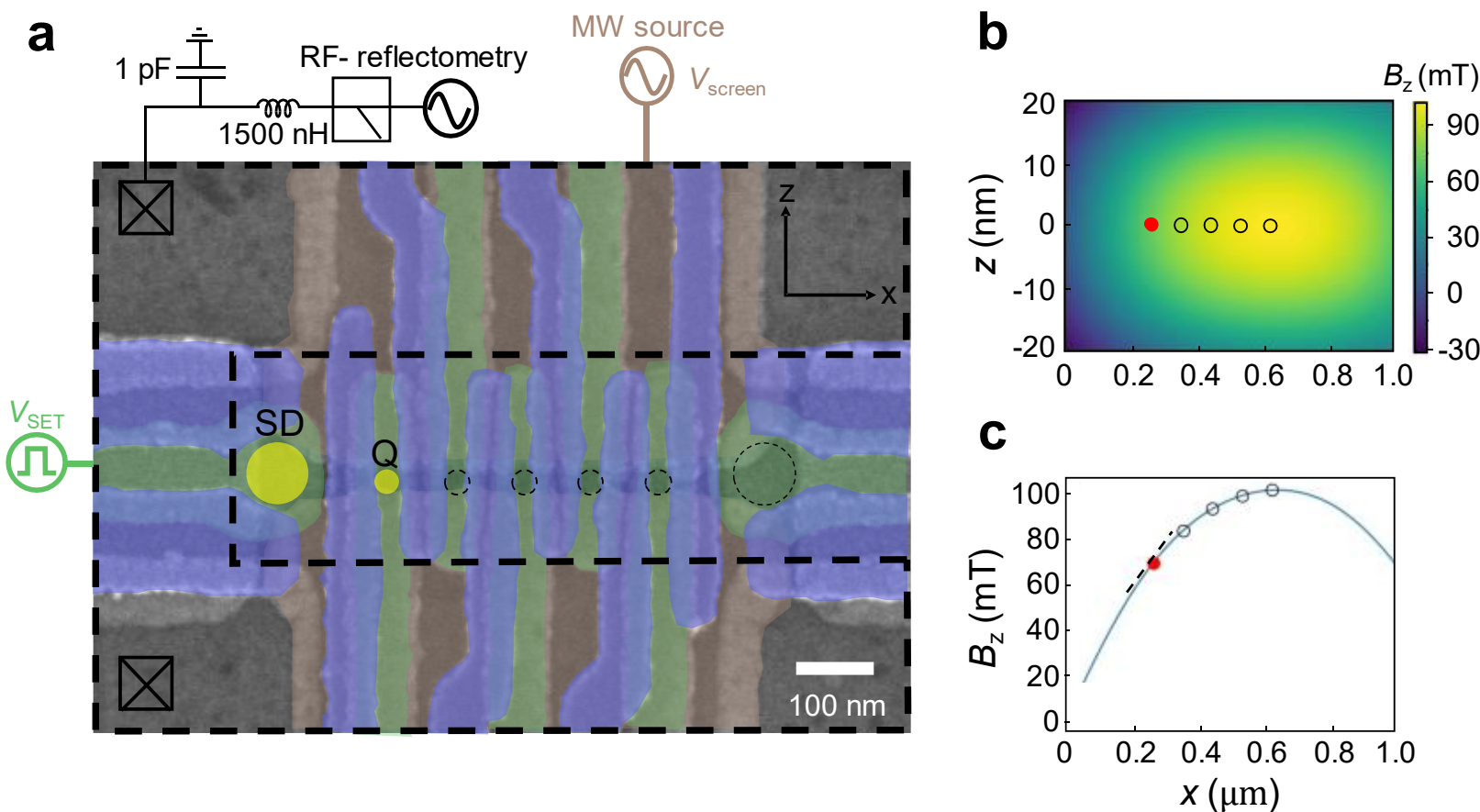

**Figure 1. Linear array $^{28}$Si/SiGe quantum dot (QD) device. a.** False-colored scanning electron microscopy image of the device before the deposition of the micromagnet. Three overlapping gate layers were fabricated on an isotopically purified $^{28}$Si/SiGe heterostructure. The first (brown), second (green), and third (purple) layers contain screening gates, plunger gates, and barrier gates for the qubit array and sensor dots, respectively. The QDs not used for the current experiment are indicated by dotted circles. The shape of the micromagnet is depicted by a bold dashed line. **b.** Numerical field distribution near the qubit area produced by the micromagnet. Circles indicate the expected locations of QDs. **c.** Field profile along the qubit array axis (z = 0 nm). Circles indicate measured qubit resonance frequencies while the solid curve is numerically simulated. The black dashed line represents the gradient of $B_z$, $dB_z/dx$ = 0.184 mT/nm at the location of the qubit. In **b** and **c**, the magnetic field values are offset by the externally applied $B_z$ = 440 mT.

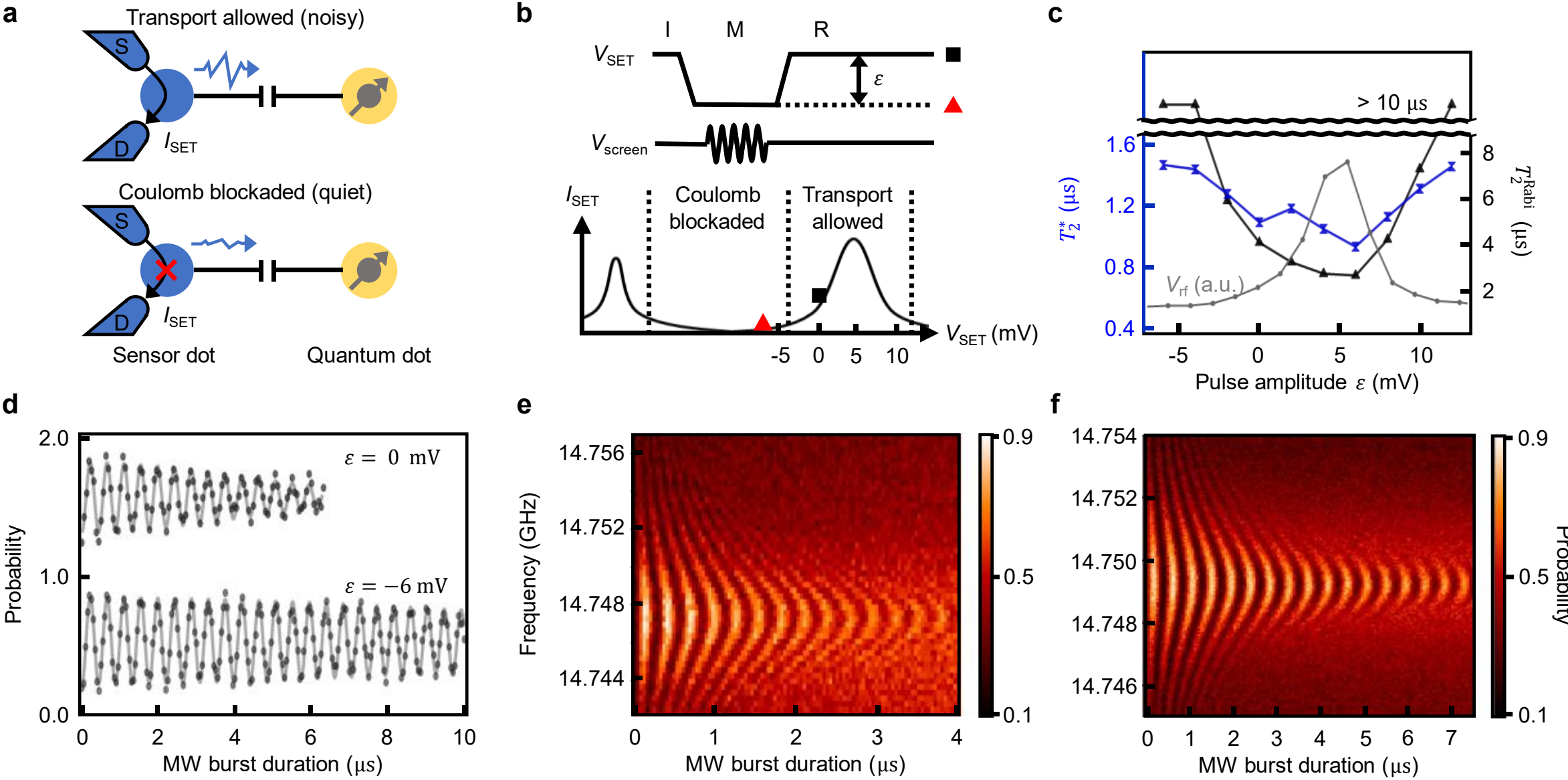

**Figure 2. Passive suppression of transduced noise. a.** Schematic diagram of the proximal charge sensor and qubit where the sensor dot (SD) is tunnel-coupled to the source and drain, and capacitively coupled to the qubit dot. Top: In the transport-allowed regime, the charge sensor is active and sensitive to the change of charge number in the qubit, but its backaction to the qubit is also enhanced. Bottom: The charge sensor in the Coulomb blockade regime disturbs the qubit minimally. **b.** Sequence for dynamic pulsing of SD chemical potential. The SD plunger gate voltage $V_{SET}$ is set at the maximally sensitive point (bottom plot, black box) during the qubit initialization (I) and readout (R). For qubit manipulation (M), $V_{SET}$ is pulsed to the Coulomb blockaded regime (bottom plot, red triangle). **c.** Inhomogeneous coherence time $T_2$* (blue) and Rabi decay time $T_2^{Rabi}$ (black) with varying dynamic pulsing voltage $\varepsilon$. The rf-reflectometry signal proportional to the transport current $I_{SET}$ is plotted in gray. The data acquisition time for each point is 5 minutes. **d.** Comparison of Rabi oscillations with and without dynamic pulsing of SD. The two oscillating probabilities are offset by 1.0 for clarity. **e, f.** Rabi chevron patterns when $\varepsilon = 0$ mV (**e**) and $\varepsilon = -6$ mV (**f**).

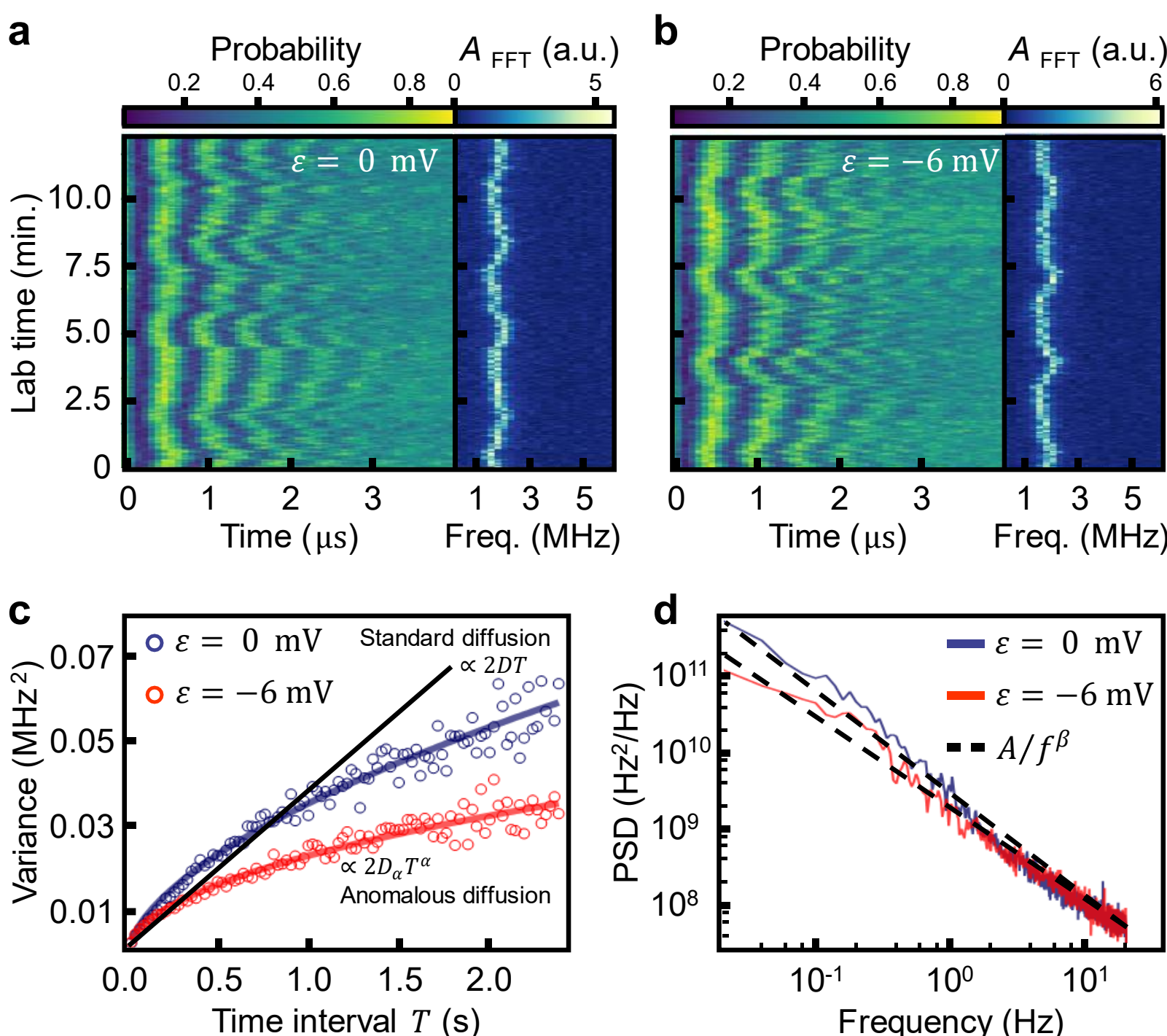

**Figure 3. Noise spectroscopy. a, b.** Repeated Ramsey interference as a function of the free evolution time with a fixed off-resonance microwave frequency for $\varepsilon = 0$ mV (**a**) and $\varepsilon = -6$ mV (**b**). Panels on the right in **a** and **b**: Fast Fourier transform (FFT) of the time domain data. **c.** Variance of the resonance frequency fluctuation as functions of the time interval $T$. The solid curves are fit to functions of the form $2D_\alpha T^\alpha$ where $D_\alpha$ is a diffusion coefficient. The linear line is overlayed as an example of standard diffusion, i.e., Brownian motion. **d.** The power spectral density PSD of the noise determined by Bayesian inference. The dashed lines are fit to the power-law decay of the form $A/f^\beta$.

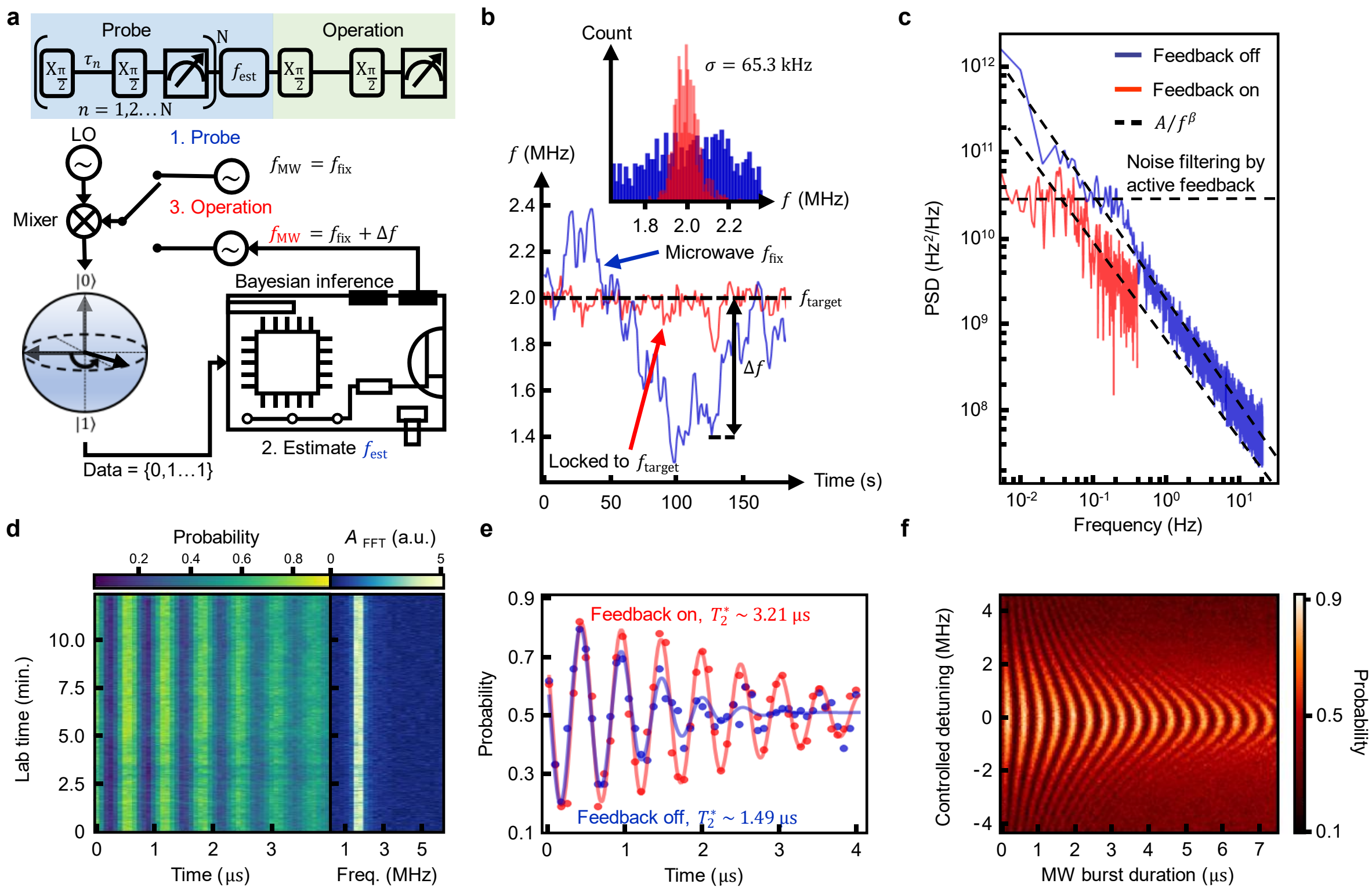

**Figure 4. Active feedback control. a.** Top: Experimental sequence of feedback control. N single-shot outcomes are collected in the probe phase, followed by estimation of the qubit frequency $f_{est}$ by the Bayesian inference circuit. The adaptive correction of the microwave frequency $f_{MW}$ is performed for the qubit operation in the next round. Bottom: Schematic block diagram for the experimental implementation of the closed-loop feedback control of the qubit frequency. **b.** Main panel: Time trace of the qubit frequencies, i.e., the oscillation frequencies of the Ramsey fringe, with (red) and without (blue) frequency locking showing frequency stabilization to the target of 2 MHz with the frequency feedback. Inset: Histogram of the qubit frequencies showing achievement of frequency uncertainty $\sigma$ = 65.3 kHz with closed-loop control. **c.** PSD of the noise with (red) and without (blue) frequency feedback. The dashed lines represent a power-law fit, yielding the noise amplitude $A$ of 837 kHz$^2$/Hz (1553 kHz$^2$/Hz) and the exponent $\beta$ of 0.945 (1.176) with (without) the frequency feedback. **d.** Coherent Ramsey experiments with feedback control showing a stabilized oscillation frequency of about 1.7 MHz.

**e.** Comparison of $T_2$* measured by Ramsey experiments with (red) and without (blue) frequency feedback. The coherence time is measured by fitting the experimental data to the Gaussian decay (solid curves). **f.** Improved quality of coherent Rabi chevron pattern in adaptive frequency control mode.

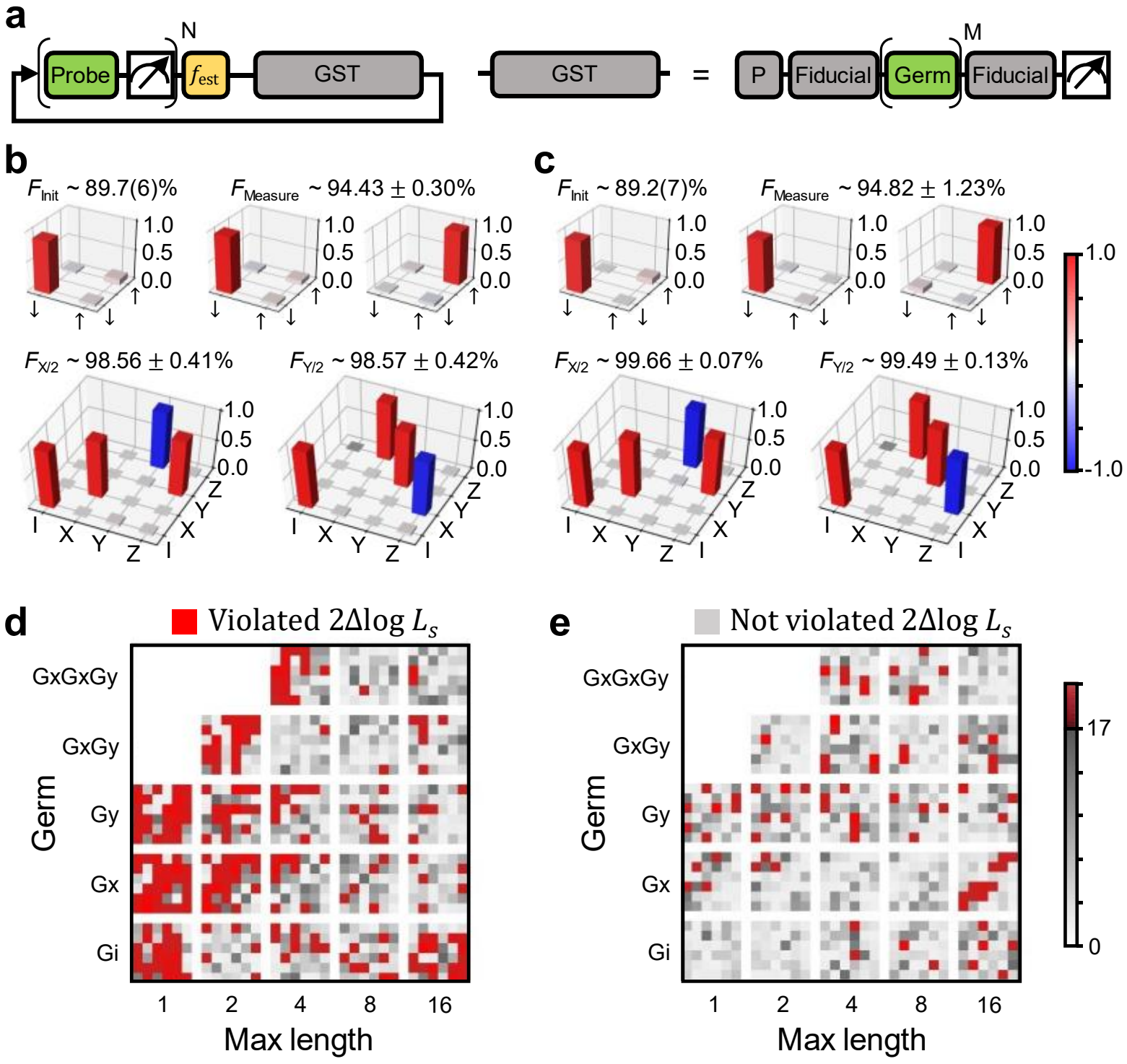

**Figure 5. Gate set tomography and model violation. a.** A gate set tomography (GST) sequence combined with active frequency feedback. **b, c.** Top row: Density matrices obtained by GST showing fidelity of state preparation and measurement without (**b**) and with (**c**) active and passive noise suppression techniques. Bottom row: Pauli transfer matrices of logic gates without (**b**) and with (**c**) active and passive noise suppression techniques. Error bars represent 95% ($2\sigma$) confidence intervals computed using Monte Carlo bootstrap resampling[4,7]. **d, e.** Model violation plot without (**d**) and with (**e**) active and passive noise suppression. The maximum length is the number of gate operations in a germ circuit. The red marks indicate

detections of model violation at a confidence level of more than 95% and the gray boxes indicate statistical fluctuations.

# Passive and active suppression of transduced noise in silicon spin qubits

Jaemin Park[1†], Hyeongyu Jang[1†], Hanseo Sohn[1], Jonginn Yun[1], Younguk Song[1], Byungwoo Kang[1], Lucas E. A. Stehouwer[2], Davide Degli Esposti[2], Giordano Scappucci[2], and Dohun Kim[1*]

*[1]Department of Physics and Astronomy, and Institute of Applied Physics, Seoul National University, Seoul 08826, Korea*

*[2] QuTech and Kavli Institute of Nanoscience, Delft University of Technology, PO Box 5046, 2600 GA Delft, The Netherlands*

*[†]These authors contributed equally to this work.*

**Corresponding author: dohunkim@snu.ac.kr*

**Supplementary Information**

**Supplementary Note 1. Experimental setup**

The device was cooled in a cryogen-free dilution refrigerator (Oxford Instruments Triton-500) with a base temperature of ~7 mK. Stable dc-voltages generated by dc-sources (Stanford Research Systems, SIM928) were combined with rapid voltage pulses from an arbitrary waveform generator (Zurich Instruments, HDAWG) via bias tees and applied to the gate electrodes. An onboard inductor of 1500 nH and a parasitic capacitance on the order of 1 pF formed a LC-tank circuit with a resonance frequency at 143 MHz for charge sensing. The reflected carrier signal was initially amplified by 45 dB with the cryogenic amplifier (Caltech Microwave Research Group, CITLF2 x2 in series) at the 4 K plate, and then additionally amplified by 25 dB at room temperature using a custom-built rf amplifier. After the reflected carrier signal was demodulated by a lock-in amplifier (Zurich Instruments, UHFLI), the signal was collected by the quantum controller (Quantum Machines, Operator-X+) at a rate of 1

MSa/s. We used Octave by Quantum Machines combined with Operator-X+ for generation of vector modulated microwave bursts, which are applied to the gate electrode $V_{screen}$. The Quantum Universal Assembly (QUA) language framework by Quantum Machines was used for overall timing control, job scheduling, defining pulse sequences for manipulation, real-time Bayesian inference, and adaptive adjustments of the frequency detuning.

**Supplementary Note 2. Details of sensor-qubit coupling**

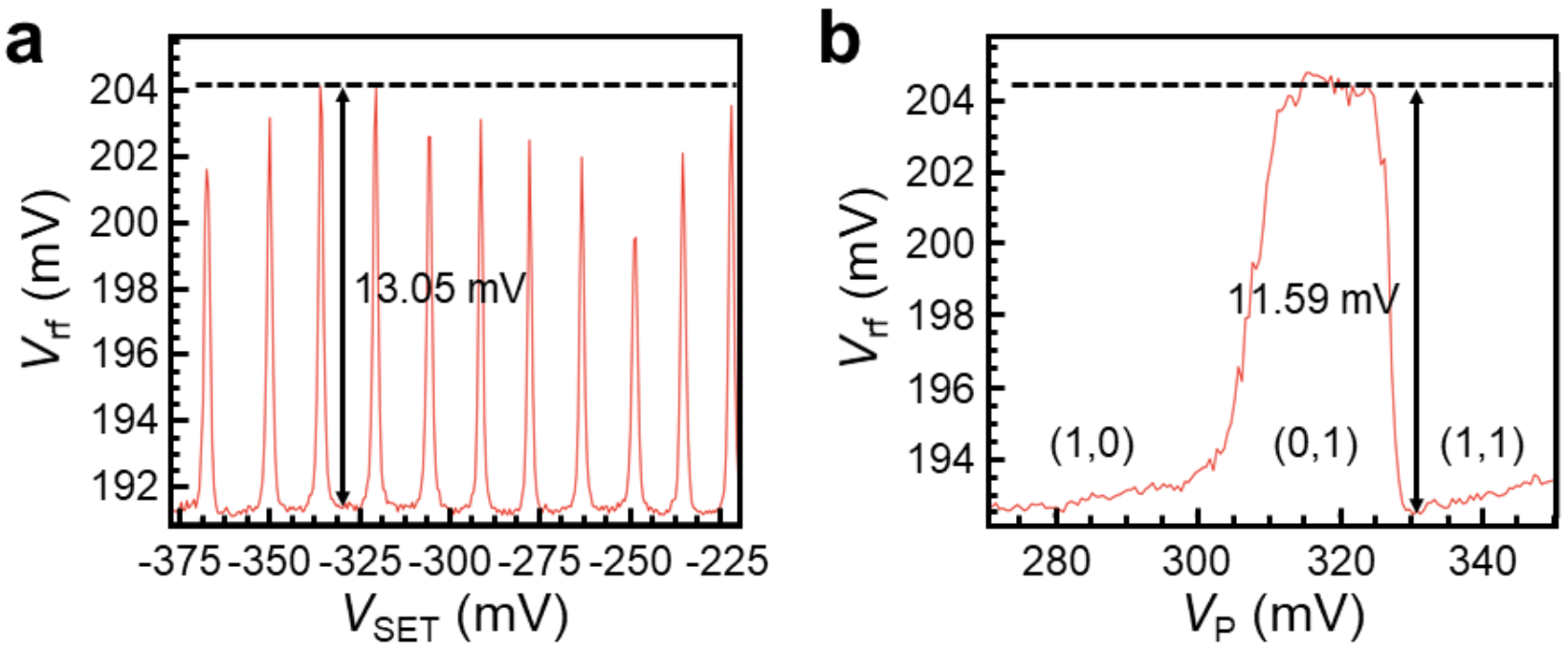

**Supplementary Figure 1. Coulomb oscillation and charge sensing. a.** Coulomb oscillation as a function of sensor plunger gate voltage. **b.** Charge sensing of the double quantum dot (DQD), as a function of QD plunger gate voltage. (n,m) indicates the charge configuration of DQD with the number of electrons n (m) occupying the left (right) QD.

The signal contrast according to the change in the number of electrons, which depends on the device design, is crucial for discriminating the spin state by spin-to-charge conversion. In this context, our device exhibited relatively strong coupling between the sensor dot (SD) and the quantum dot (QD), allowing for sensitive charge detection. In Supplementary Fig. 1a and 1b, we observed a signal of one electron tunneling of 11.59 mV which is close to the peak-to-peak amplitude of the coulomb oscillation of 13.05 mV indicative of strong capacitive coupling between the sensor and QD. We suppose that the strong coupling as well as the linear array structure parallel to the direction of the large magnetic field gradient can induce significant backaction on the qubit.

**Supplementary Note 3. Details of noise spectroscopy**

In a 24 ms time interval step *T*, we extracted frequency increments from the estimated qubit frequency, such as $dX_T = X_{t+T} - X_t$, where $dX_T$ is the increment during *T* and $X_t$ is the qubit frequency estimated at *t*. Subsequently, we calculated the variance of $dX_T$, referred to as $\sigma^2 = 2D_\alpha T^\alpha$, which represents the mean-squared displacement the state randomly jumps (random walk) to the next state during *T*. To compute the noise power spectral density (PSD), we applied a periodogram method to analyze the frequency components of the noise signal. The periodogram method differs from the Bayesian approach for computing the PSD[1]. The frequentist approach, such as the periodogram, provides computed values at each point, whereas the Bayesian approach provides probability distribution functions at each point.

To explain the decoherence function during the free evolution, we considered random phase accumulation with the free evolution time *t*

$$\phi(t) = \int_{-\infty}^{\infty} dt' 2\pi \nu(t') \theta_t(t') \quad (1)$$

where the noise in qubit frequency $\nu(t')$ follows a Gaussian distribution and $\theta_t(t')$ is 1 (0) when the qubit is evolved (not evolved) in time. If the noise is a wide-sense stationary (WSS) process, the autocorrelation function only depends on $\tau = t' - t''$ such that $R_{\nu\nu}(\tau)$ = $\langle \nu(t')\nu(t'') \rangle$. Using the Wiener-Khinchin theorem, we can calculate the mean-squared phase noise

$$\begin{aligned}\left\langle \phi^2(t)\right\rangle &= (2\pi)^2\left\langle \int_0^t dt'\nu(t')\int_0^t dt''\nu(t'')\right\rangle \\ &= (2\pi)^2\int_0^t\int_0^t dt'dt''\int_{-\infty}^{\infty} df\, S(f)\, e^{2\pi i f(t'-t'')} \qquad (2)\\ &= (2\pi)^2\int_{-\infty}^{\infty} df\, S(f)\, F_t(f)\end{aligned}$$

where $S(f) = A/f^{\beta}$ is the power spectral density and $F_t(f)$ is the filter function[2] given by

$$F_t(f) = \left|\int_0^t dt' e^{2\pi i f t'}\right|^2 = t^2\,\mathrm{sinc}^2(\pi f t) \qquad (3)$$

For comparison, the filter function for spin echo experiments is an integral from 0 to $t/2$ ($t/2$ to $t$) with $\theta_t(t')$ = 1 (-1), accounting for spin refocusing. Using Eqns. (2) and (3), we can find the decoherence function

$$W(t) = \exp\left(-\left\langle \phi^2(t)\right\rangle \Big/ 2\right) = \exp\left(-\frac{t^2}{2}(2\pi)^2\int_{f_0}^{\infty} df\, S(f)\,\mathrm{sinc}^2(\pi f t)\right) \qquad (4)$$

In a quasi-static approximation ($f << 1/t$), we can let $\mathrm{sinc}^2(\pi f t)$ ~ 1. The integral in Eqn. (4) becomes

$$\begin{aligned}W(t) &= \exp\left(-\frac{t^2}{2}(2\pi)^2 A\int_{f_0}^{f_1} df\, f^{-\beta}\right)\\ &= \exp\left(-\frac{t^2}{2}(2\pi)^2\sigma_{\text{static}}^2\right) = \exp\left(-t^2\Big/\left(T_2^*\right)^2\right) \qquad (5)\end{aligned}$$

where $f_1$ of 0.1 MHz is the quasi-static limit determined by $t$, which ranges from a few nanoseconds to microseconds, and $f_0$ is the reciprocal of the total integration time of the experiment. If the quasi-static approximation is not valid, the decoherence function would be no longer the Gaussian decay function. However, our estimation rate of 41.6 Hz properly lies within the quasi-static limit, enabling noise analysis based on Hamiltonian parameter estimation.

## Supplementary Note 4. Qubit frequency estimation in the presence of exchange interactions

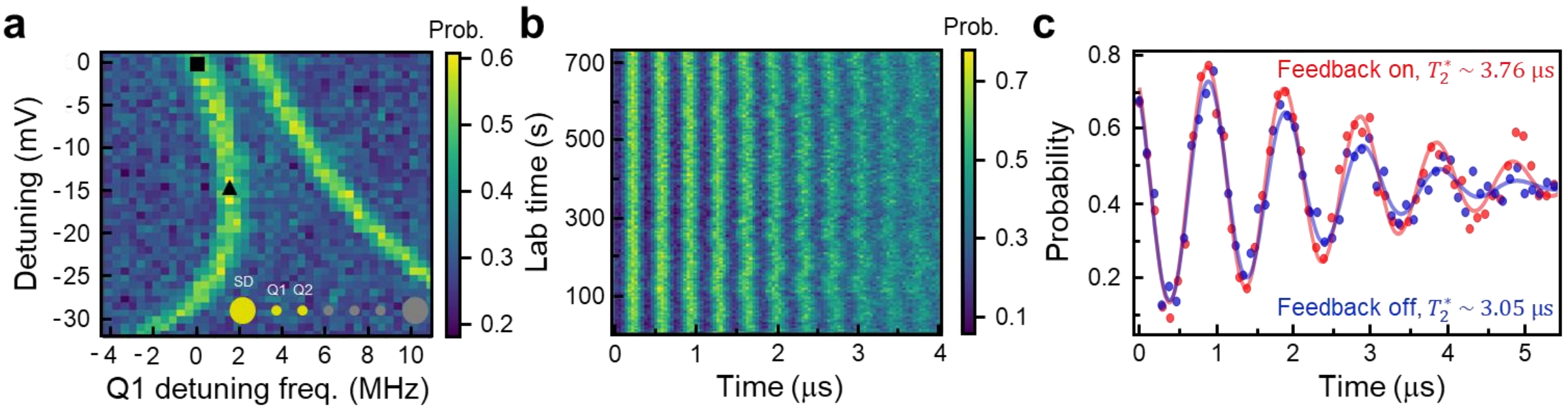

**Supplementary Figure 2. Exchange spectroscopy and active suppression of noise. a.** Detuning frequency and voltage are varied to obtain electric dipole spin resonance (EDSR) spectra of Q1 in the exchange-always-on DQD. The positions of Q1 and Q2 are indicated in the bottom right. **b.** Repeated Ramsey interference of Q1 as a function of free evolution time at the detuning = –15 mV, as depicted by the black triangle in **a**, while active feedback control is not used. **c.** Comparison of inhomogeneous coherence time $T_2$* measured by Ramsey experiments with (red) and without (blue) frequency feedback. The coherence time is measured by fitting the experimental data to the Gaussian decay (solid curves).

We investigated the exchange coupling[3] by applying a $\pi/2$ pulse to Q2 to implement conditional rotation of Q1. Supplementary Figure 2a shows that the increasing exchange energy contributes to the nonlinear geometry of Q1 resonance frequency, while the linear contribution of the micromagnet is observed around zero detuning where the exchange energy is minimized. At the detuning = –15 mV, the qubit system could be more susceptible to the exchange noise, but simultaneously robust to the slow dephasing noise as demonstrated in Supplementary Fig. 2b. Consequently, Supplementary Fig. 2c shows a modest improvement in $T_2$* from 3.05 to 3.76 μs, achieved by active feedback control. The result shows that the developed active feedback protocol can also be used when two-qubit interaction is turned on where repeated Bayesian estimation of the target qubit with different control qubit states can efficiently result in full two-qubit parameter estimation including exchange energy.

**Supplementary Note 5. The potential presence of high-frequency noise**

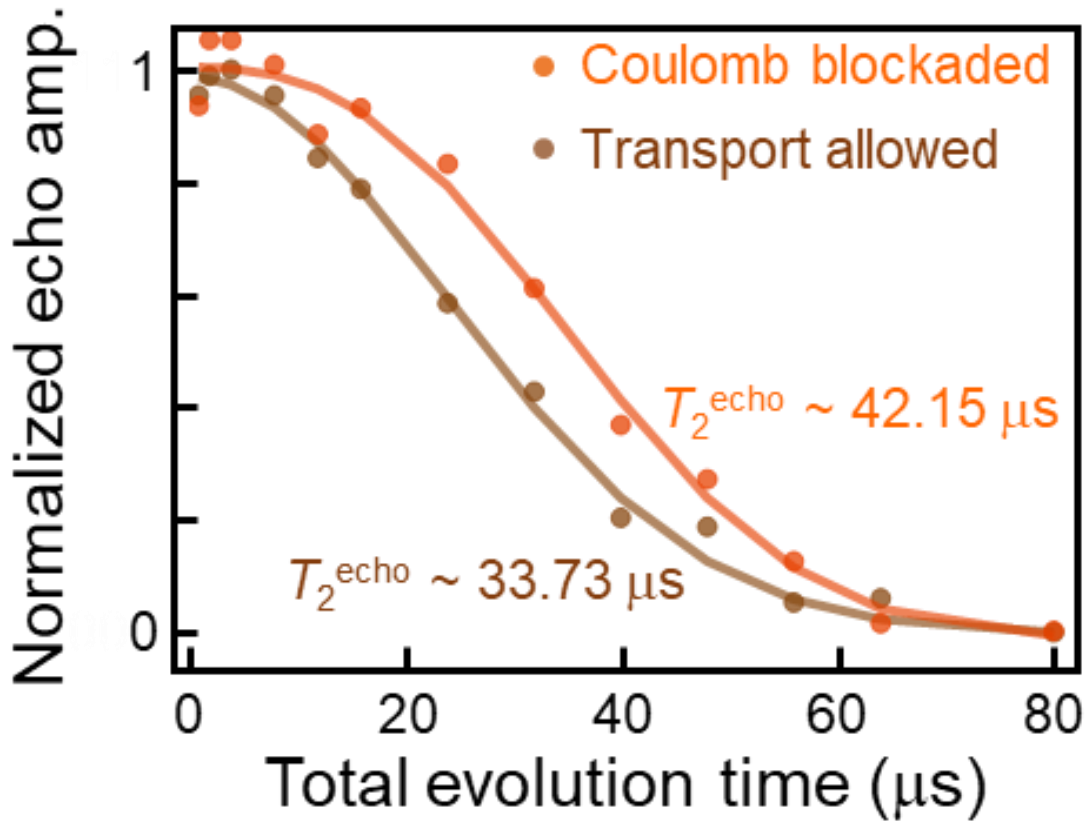

**Supplementary Figure 3. Spin echo experiments.** Echo signal as a function of total evolution time $\tau$, with (orange) and without (brown) the passive technique. Each data point is acquired with the phase-varying spin echo pulse ($\pi$/2)-($\tau$/2 waits)-($\pi$)-($\tau$/2 waits)-($\pi/2)_\theta$, representing the oscillation amplitude by sweeping the axis of the final $\pi$/2 rotation. The solid lines are fit to the form of $\exp(-(\tau/T_2^{echo})^{\alpha})$, yielding $\alpha = 2$ (brown) and $\alpha = 2.58$ (orange).

From the noise trajectories obtained from Bayesian inference, low-frequency noise characteristics were analyzed with and without the passive technique. However, due to the limited bandwidth of a few tens of Hz, this method could not capture high-frequency noise characteristics. To investigate the presence of high-frequency noise, possibly caused by the SD transport current, we performed spin echo experiments using the passive technique. We observed an increase in $T_2^{echo}$ from 33.73 to 42.15 μs when the passive technique was applied, as shown in Supplementary Figure 3, indicating that the passive technique suppresses high-frequency noise which the spin echo technique could not address. Thus, the improvement of $T_2^{echo}$ strongly suggests that the original sensor noise contains a significant portion of high-frequency noise. Although our approach did not focus on identifying the exact nature of the sensor noise responsible for the high-frequency noise spectrum, it deserves further studies. For better understanding, one can conduct more refined experiments including high-order dynamic decoupling, such as Carr-Purcell-Meiboom-Gill pulse sequences, Rabi spectroscopy, and spin-locking sequences.

**Supplementary Note 6. Additional information on the coherence time**

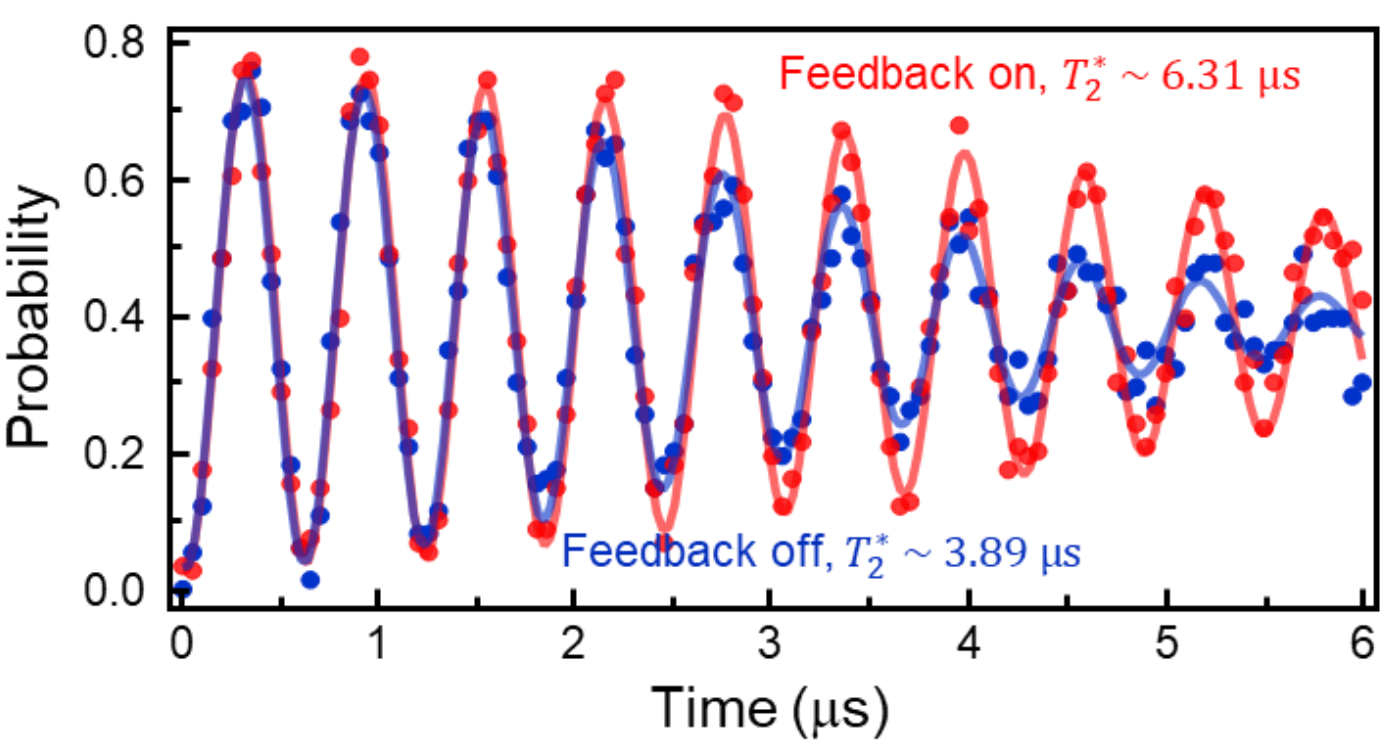

**Supplementary Figure 4. Comparison of the coherence time.** $T_2$* measured by Ramsey experiments with (red) and without (blue) frequency feedback. The data were acquired using a Pauli spin blockade-based parity measurement[4,5] with a reduced estimation cycle from 24 to 10 ms as opposed to the EST readout method used in the main text, and in another device which has the same design as the one used in the main text. Noting that feedback-off $T_2$* of 3.89 μs already exceeds the enhanced $T_2$* of 3.21 μs in the main text, an increase of $T_2$* to 6.31 μs shows the potential of active feedback control based on noise estimation.

**Supplementary Note 7. Impact of active noise suppression alone**

In the main text, we investigated the effect of active feedback control when the passive noise suppression technique was applied by default. However, even with active noise suppression alone, we confirmed that $T_2$* increased by 250% from 1.13 to 2.83 μs. Although it remained below the 3.21 μs achieved with the 'passive + active' approach in the main text, the difference in $T_2$* between the 'active only' and 'passive + active' approaches implies that high-frequency noise, which could be suppressed by the passive technique (see Supplementary

Figure 3), has less impact during free evolution.

In contrast, for Rabi experiments, improvements in driving quality were not significant and difficult to observe when using only active noise suppression, possibly due to the sensor noise being a primary factor limiting $T_2^{\mathrm{Rabi}}$. Thus, to significantly improve both $T_2^*$ and $T_2^{\mathrm{Rabi}}$, we applied the passive noise suppression technique by default and used the combined approach.

## Supplementary Note 8. Additional information on model violation

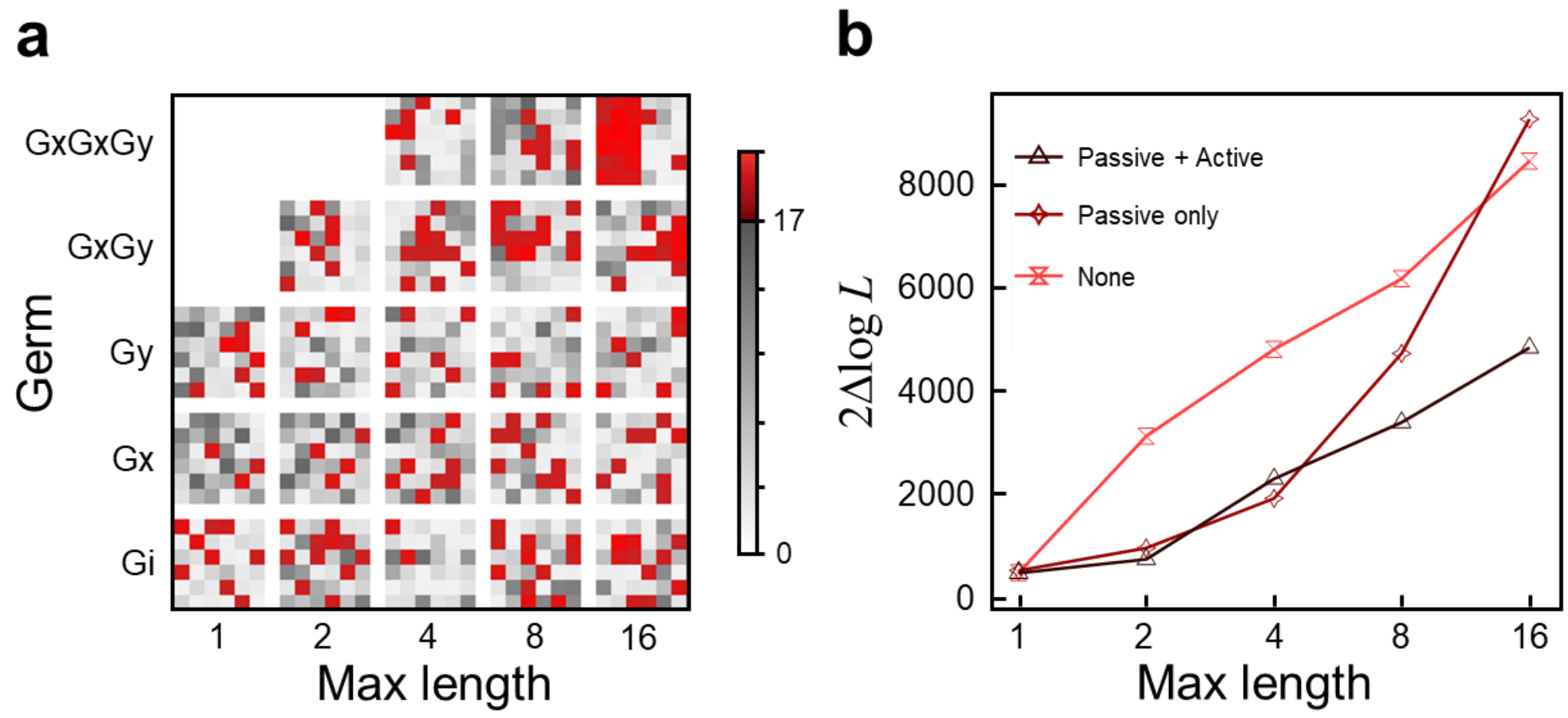

**Supplementary Figure 5. Additional information on model violation. a.** Model violation plot using only the passive technique. The fidelities $F_{\mathrm{init}}$, $F_{\mathrm{measure}}$, $F_{\mathrm{X/2}}$ and $F_{\mathrm{Y/2}}$ are 82.9 ± 0.97%, 88.57 ± 1.49%, 99.36 ± 0.47% and 99.35 ± 0.55% respectively (data not shown). **b.** Comparison of three cases by evaluating $2\Delta\log L$. The 'Passive only' data is likely to follow the data obtained using both techniques up to a maximum length of 4, after which it approaches the data without any techniques as the maximum length increases.

## Supplementary References